\documentclass[12pt,preprint]{aastex}

\usepackage{mathrsfs}
\usepackage{amsmath}
\usepackage{amssymb}
\usepackage{graphicx}

\newcommand{\msl}{\mathscr{L}}

\begin{document}

\title{New low-frequency nonlinear electromagnetic wave in a magnetized plasma}

\author{M.\ Marklund, P.K.\ Shukla, L.\ Stenflo, G.\ Brodin, and M.\ Servin}
\affil{Department of Physics, Ume{\aa} University, SE--901 87 Ume{\aa}, Sweden}


\begin{abstract}
A new nonlinear electromagnetic mode in a magnetized plasma is
predicted. Its existence depends on the interaction of an intense circularly 
polarized electromagnetic wave with a plasma, where quantum
electrodynamical photon--photon scattering is taken into account. This
scattering gives rise to a new coupling between the matter and the
radiation.  Specifically, we consider an electron--positron plasma, and show 
that the propagation of the new mode is admitted. It could
be of significance in pulsar magnetospheres, and result in 
energy transport between the pulsar poles. 
\end{abstract}

\maketitle

Astrophysical environments can be most violent and energetic. 
Physics considered `exotic' in Earth based laboratory applications can 
be common throughout our Universe, and sometimes even vital for the existence 
of certain observed phenomena. Pulsars, surrounded by strong magnetic fields, 
are most prolific sources of exotic physics. Quantum electrodynamics (QED) is 
an indispensable explanatory model for much of the observed pulsar phenomena. 
Scattering of photons off photons is predicted by QED, and it can be
a prominent component of pulsar physics, since pulsars offer 
the necessary energy scales for such scattering to occur. Related
to the scattering of photons is the concept of photon splitting in 
strong magnetic fields (Adler 1971). It has been suggested that such effects 
could be important in explaining the radio silence of magnetars 
(Kouveliotou 1998; Baring \& Harding 2001). In the present Letter we will
point out the existence of a new low-frequency nonlinear electromagnetic wave 
that may exist in a magnetized plasma, due to the interaction of photons with 
the quantum vacuum. A discussion of the properties of this new electromagnetic wave
using parameters relevant to strongly magnetized pulsars will be given.

The weak field theory of photon--photon scattering can be formulated in terms 
of the effective Lagrangian density

\begin{equation}
  \msl = \msl_0 + \msl_{\text{HE}},
\end{equation}
where $\msl_0 = -\tfrac{1}{4}\epsilon_0F_{ab}F^{ab} =
\tfrac{1}{2}\epsilon_0({\bf E}^2 - c^2{\bf B}^2)$ is the classical free
field Lagrangian, and 

\begin{equation}\label{eq:lagrangian}
  \msl_{\text{HE}} = \kappa\epsilon_0^2\left[4\left(
  \tfrac{1}{4}F_{ab}F^{ab}\right)^2 +
  7\left( \tfrac{1}{4}F_{ab}\widehat{F}^{ab} \right)^2 \right],
\end{equation}
is the Heisenberg--Euler correction (Heisenberg \& Euler 1936; 
Schwinger 1951), where $\widehat{F}_{ab} = \tfrac{1}{2}\epsilon_{abcd}F^{cd}$, 
and $\tfrac{1}{4}\widehat{F}_{ab}F^{ab} = - c{\bf E}\cdot{\bf B}$.
Here $\kappa \equiv 2\alpha^2\hbar^3/45m_e^4c^5 \approx 1.63\times
10^{-30}\, \mathrm{m}\mathrm{s}^{2}/\mathrm{kg}$, $\alpha$ is
the fine-structure constant, $\hbar$ is the Planck constant, $m_e$ is the 
electron mass, and $c$ is the speed of light in vacuum.
With $F_{ab} = \partial_aA_b - \partial_bA_a$, $A^b$ being the four-potential, 
we obtain, from the Euler--Lagrange equations, the field equations  
$\partial_b[\partial\msl/\partial F_{ab}] = 0$, i.e.\ (see, e.g.\ Shukla \textit{et al.}\ 2004)

\begin{equation}
  \partial_bF^{ab} = 2\epsilon_0\kappa\partial_b\left[ 
  (F_{cd}F^{cd})F^{ab} 
  + \tfrac{7}{4}(F_{cd}\widehat{F}^{cd})\widehat{F}^{ab} \right]
  + \mu_0 j^a ,
  \label{eq:maxwell}
\end{equation}
where $j^a$ is the four current.

For a circularly polarized wave $\mathbf{E}_0 = E_{0} 
(\hat{\mathbf{x}} \pm i\hat{\mathbf{y}})\exp(i\mathbf{k}\cdot\mathbf{x}
  - i\omega t)$ propagating along a constant magnetic field ${\bf B}_0 = 
B_{0}\hat{\mathbf{z}}$, the invariants satisfy

\begin{equation} 
  F_{cd}F^{cd} = -2E_0^2\left( 1 - \frac{k^2c^2}{\omega^2}\right) 
    + 2c^2B_0^2 
  \, \text{ and } \,
  F_{cd}\widehat{F}^{cd} = 0 ,
\end{equation}
where $k$ is the wave number and $\omega$ the frequency of the
circularly polarized electromagnetic wave. Thus, Eq.\ (\ref{eq:maxwell}) can be written as

\begin{equation}
  \Box A^a = -4\epsilon_0\kappa\left[ E_0^2\left( 1 - \frac{k^2c^2}{\omega^2}\right) 
    - c^2B_0^2 \right]\Box A^a + \mu_0 j^a ,
  \label{eq:wave}
\end{equation}
in the Lorentz gauge, and $\Box = \partial_a\partial^a$.  For circularly polarized 
electromagnetic waves propagating in a magnetized cold multicomponent plasma, the four 
current can be `absorbed' in the wave operator on the left-hand side by the replacement

\begin{equation}
  \Box \rightarrow -D(\omega, k) ,
\end{equation} 
where $D$ is the plasma dispersion function, given by 
(see, e.g. Stenflo (1976) and Stenflo \& Tsintsadze (1979))

\begin{equation}
  D(\omega,k) = k^2c^2 - \omega^2 + 
  \sum_j\frac{\omega\omega_{pj}^2}{\omega\gamma_j \pm \omega_{cj}} ,
\label{eq:dispersionfunction}   
\end{equation}
where the sum is over the plasma particle species $j$,  

\begin{equation}
  \omega_{cj} = \frac{q_jB_0}{m_{0j}} \quad \text{ and } \quad
  \omega_{pj} =
  \left(\frac{n_{0j} q_j^2}{\epsilon_0 m_{0j}}\right)^{1/2} ,
\label{eq:defs}
\end{equation}
is the gyrofrequency and plasma frequency, respectively, and 

\begin{equation}
  \gamma_j = (1 + \nu^{2}_j)^{1/2}, 
\end{equation}
is the the gamma factor of species $j$, with $\nu_j$ satisfying

\begin{equation}
  \nu^{2}_j = \left(
  \frac{eE_0}{cm_{0j}} \right)^2\frac{1 + \nu^{2}_j}{[\omega(1 +
  \nu^{2}_j)^{1/2} \pm \omega_{cj}]^2} .
  \label{eq:nu}
\end{equation}
Here $n_{0j}$ denotes particle density in the laboratory frame and $m_{0j}$ particle rest mass. 
It should be emphasized that for the case of circularly polarized waves propagating 
along an external magnetic field, the effects due to the electron and ion currents can be 
calculated without linearizing the plasma governing equations (see e.g.  Derby 1978, Goldstein 1978, 
Stenflo \& Tsintsadze  1979 and Stenflo \& Shukla 2001).  Thus, in Eq.\ (\ref{eq:dispersionfunction}), 
the relativistic nonlinearity, the full Lorentz force as well as the other plasma  nonlinearities are 
fully accounted for.  The only limiting assumption of the electromagnetic field amplitude is due to 
the validity of the Euler-Heisenberg Lagrangian. We note that the application of Eq. (2) requires field 
strengths smaller than the Schwinger critical field $E_S = m_e^2c^3/e\hbar \sim 10^{18}$ V/m 
(see e.g. Schwinger 1951). The dispersion relation, obtained from Eq.\ (\ref{eq:wave}), then reads

\begin{eqnarray}
  && D(\omega,k) = \frac{4\alpha}{45\pi}(\omega^2 - k^2c^2)
  \nonumber \\ &&\qquad \qquad
  \times\left[\left( \frac{E_0}{E_S}
  \right)^2\frac{\omega^2 - k^2c^2}{\omega^2} - \left(\frac{cB_0}{E_S}\right)^2 
  \right] .
\label{eq:qeddisp}
\end{eqnarray}
We note that as the plasma density goes to zero, the effect due to photon--photon scattering, 
as given by the right-hand side of Eq.\ (\ref{eq:qeddisp}), vanishes, since then 
$\omega^2 - k^2c^2 = 0$.

It may be instructive to first consider the small amplitude limit of Eq.\ (\ref{eq:qeddisp}).  
In a magnetized pair-plasma this reads

\begin{equation}
\frac{k^2c^2}{\omega^2} =1-\frac{2\omega_p^2}{\omega^2 -\omega_c^2},
\label{eq:small-amp-limit}
\end{equation}
which shows the existence of fast and slow EM waves.  In the low-frequency long wavelength limit, 
and for $\omega_p \ll \omega_c$ (as we typically have in pulsars) we then have $\omega/k\sim c$. 
The phase speed of the slow wave will thus be close to the speed of  light.  However, for large 
amplitude waves, the situation is drastically different. As we will see below, we indeed have 
then quite new kinds of dispersive electromagnetic waves whose frequencies depend on the wave 
amplitudes, and for which the phase velocities fulfill $\omega/k\ll c$. 

Next, we focus on mode propagation in an ultra-relativistic electron--positron plasma ($\gamma_{e} \gg 1$), 
where the two species have the same number density $n_{0}$. Then Eq.\ (\ref{eq:qeddisp}) gives

\begin{eqnarray}
  \!\!\!\!\!\!\!\!\!\!\!\! && k^{2}c^{2} - \omega^{2} \pm 
	\frac{\omega\omega_{pe}^{2}}{\omega_{E}} 
	\nonumber \\ \!\!\!\!\!\!\!\!\!\!\!\! && 
	= 
	\frac{4\alpha}{45\pi}\left[ \left( \frac{E_{0}}{E_{S}} 
	\right)^{2}\frac{\omega^{2} - k^{2}c^{2}}{\omega^{2}}
	- \left( \frac{cB_0}{E_S}\right)^2 \right](\omega^2 - k^2c^2) . 
	\label{eq:epdisp}
\end{eqnarray}
Following Stenflo \& Tsintsadze (1979), we have defined $\omega_{E} = eE_{0}/cm_{0e}$. 

Looking for low-frequency modes, we now use the approximation
$\omega \ll kc$, at which Eq.\ (\ref{eq:epdisp}) gives

\begin{equation}
    \frac{k^{2} c^{2} }{\omega^{2} } \approx \frac{4\alpha}{45\pi}\left[\left( 
    \frac{E_0}{E_S}  \right)^2 \frac{k^{2} c^{2} }{\omega^{2} } 
    + \left(\frac{cB_0}{E_S}\right)^2 \right]\frac{k^2c^2}{\omega^2}  \mp 
    \frac{\omega_{pe}^{2}}{\omega\omega_{E}}  .  
\label{eq:transverse2}
\end{equation}
It is sometimes advantageous to use the relation $\omega_E = \omega_e (E_0/E_S)$, 
where $\omega_e = m_ec^2/\hbar$ is the Compton frequency, to write 
Eq.\ (\ref{eq:transverse2}) as 

\begin{equation}
   \frac{k^{2} c^{2} }{\omega^{2} } \approx \frac{4\alpha}{45\pi}\left[\left( 
    \frac{E_0}{E_S}  \right)^2 \frac{k^{2} c^{2} }{\omega^{2} } 
    + \left(\frac{cB_0}{E_S}\right)^2 \right]\frac{k^2c^2}{\omega^2}  \mp 
    \frac{\omega_{pe}^{2}}{\omega\omega_e}\frac{E_S}{E_0}  .  
\label{eq:transverse3}
\end{equation}

Using  the dispersion relation (\ref{eq:transverse2}) the group velocity 
$v_{g} \equiv d\omega/dk$ is

\begin{equation}\label{eq:group}
    v_{g} = \frac{\displaystyle{1 - \frac{4\alpha}{45\pi}\left( \frac{cB_0}{E_S} \right)^2 \pm 
    \frac{2v_p}{kc^2}\frac{\omega_{pe}^{2}}{\omega_{E}}}}%
    {\displaystyle{1 - \frac{4\alpha}{45\pi}\left( \frac{cB_0}{E_S} \right)^2 \pm 
    \frac{3v_p}{2kc^2}\frac{\omega_{pe}^{2}}{\omega_{E}}}}v_{p} ,
\end{equation}
where $v_{p} \equiv \omega/k$ is the phase velocity. 

Pulsar magnetospheres exhibit extreme field strengths in a highly energetic
pair plasma. Ordinary neutron stars have surface magnetic field strengths of 
the order of $10^{6}-10^{9}\,\mathrm{T}$, while magnetars can reach 
$10^{10}-10^{11}\,\mathrm{T}$ (Kouveliotou 1998), coming close to,
or even surpassing, energy densities $\epsilon_0E_S^2$ 
corresponding to the Schwinger limit. Such strong fields will make the  
vacuum fully nonlinear, due to the excitation of virtual pairs. 
Photon splitting can therefore play a significant role in these extreme 
systems (Harding 1991; Baring \& Harding 2001). The emission of short wavelength 
photons due to the acceleration of plasma particles close to the polar caps
results in production of electrons and positrons as the photons propagate 
through the pulsar intense magnetic field (Beskin 1993). Given the 
Goldreich--Julian density $n_{GJ} = 7\times 10^{15} (0.1\,
\mathrm{s}/P)(B/10^8\,\mathrm{T})$ m$^{-3}$,
where $P$ is the pulsar period and $B$ the pulsar magnetic field, the
pair plasma density is expected to satisfy $n_0 = M n_{GJ}$, where $M$ is
the multiplicity (Beskin 1993; Luo \textit{et al.}\ 2002). Moderate estimates 
give $M = 10$ (Luo \textit{et al.}\ 2002). Thus, the density in a pulsar pair plasma 
can be of the order $10^{18}$ m$^{-3}$. The plasma experiences a relativistic factor 
$\sim 10^2 - 10^3$ (Asseo 2003). On the other hand, the primary beam
will have $n_0 \sim n_{GJ}$ and $\gamma \sim 10^6 -10^7$ (Asseo 2003).

Furthermore, we note that the pulsar magnetosphere is very far from thermodynamical equilibrium, and 
consequently there exist numerous possible excitation mechanisms. In addition to Cherenkov excitation 
and cyclotron excitation that may excite Alfven wave like modes (e.g. Lyutikov 1999), free energy sources 
in the form of temperature anisotropies and/or electron beams can drive Weibel instabilities and beam type 
instabilities, respectively.  Several of these initial excitation mechanisms can lead to
the existence of arbitrarily large amplitude waves. 

The two QED terms in the squared bracket of Eq.(\ref{eq:transverse3}) correspond to 
the wave field and the static background field respectively. The former dominates if the oscillating 
(wave) part of the magnetic field is larger than the static part, and the latter dominates if the opposite 
is true. Due to the dipole nature of the static field, the former case is more likely to hold at a 
comparatively large distance from the pulsar surface, whereas close to the surface the static part is 
likely to dominate.  From now on we concentrate on comparatively weak background fields strengths 
(thus excluding regions in the immediate vicinity of the pulsar surface). Accordingly we have 
$cB_{0}\ll E_{S}$, and we therefore drop the term proportional to $B_{0}^{2}$ in 
Eq.\ (\ref{eq:transverse3}).  Next, using the normalized quantities 
$\Omega = \omega \omega _{e}/\omega _{pe}^{2}$, $K=(4\alpha /45\pi
)^{-1/2}kc\omega _{e}/\omega _{pe}^{2}$ and $\mathcal{E}=(4\alpha /45\pi)E_{0}/E_{S}$, 
the dispersion relation (\ref{eq:transverse3}) reads 

\begin{equation} \label{eq:norm}
\Omega^{2} = \mathcal{E}^{2}K^{2}\mp \frac{\Omega^{3}}{\mathcal{E}K^{2}} .
\end{equation}
This dispersion relation describe three different modes, two with 
$+$ polarization and one with $-$ polarization. The normalized frequency as 
a function of $K$ and $\mathcal{E}$ is shown in Fig.\ 1.  
We note that for $K \ll 1$, the dispersion relation (\ref{eq:norm}) agrees with
that of Stenflo \&Tsintsadze (1979), whereas in the opposite limit $K \gg 1$, 
the QED term in (\ref{eq:norm}) is dominating. 
For the given density, the latter regime applies, except for
extremely long wavelengths ($>10^{8}\,\mathrm{m}$), and thus we note 
that QED effects are highly relevant for the propagation of these modes in 
the pulsar environment. For small $K$ there is
only one mode, but as seen from the second and third panels of Fig.\ 1, 
two new modes appear for $K \gtrsim 2.6$. Thus for large $K$, applicable in the pulsar
environment, there are three low-frequency modes ($\omega \ll kc$) that
depend on nonlinear QED effects for their existence.  Using $cB_0 \ll E_S$, 
the expression (\ref{eq:group}) for the group velocity becomes  

\begin{equation}
  \frac{d\Omega}{dK} =\frac{\Omega \pm 2\Omega^2/\mathcal{E}K^2}{\Omega \pm
  3\Omega^2/2\mathcal{E}K^2}\frac{\Omega}{K},
\label{eq:group-vel}
\end{equation}
and thus we see that the propagation speed depends nonlinearly on the plasma parameters.  
Furthermore, from (\ref{eq:group-vel}), and recalling the normalizations used, we 
see that the group-velocity is typically well below the speed of light.  We emphasize that the 
low group velocity by itself makes our mode more long-living, since the energy supplied to the 
mode in order to balance the convective energy loss is proportional 
to the group velocity.  We thus suggest that the three new modes presented above 
can contribute to an understanding of the very complicated energy transport 
phenomena taking place in the accretion discs of pulsars.

In summary, we have reported the existence of a new low-frequency nonlinear electromagnetic 
wave in a magnetized plasma. The dispersion relation of the wave has been presented, 
and analysed using relevant astrophysical parameters. Applications to pulsar 
magnetoplasmas have been pointed out.

\clearpage

\begin{figure*}[th]
\includegraphics[width=.3\textwidth]{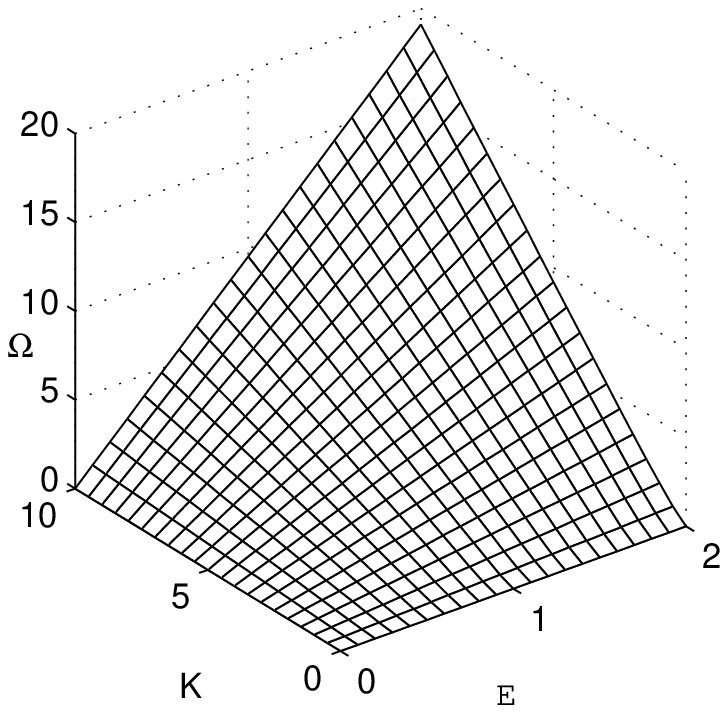} 
\includegraphics[width=.3\textwidth]{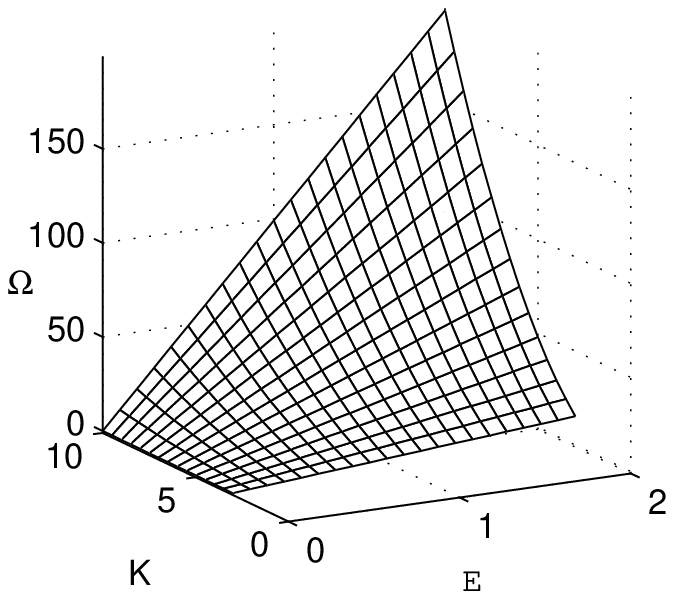} 
\includegraphics[width=.3\textwidth]{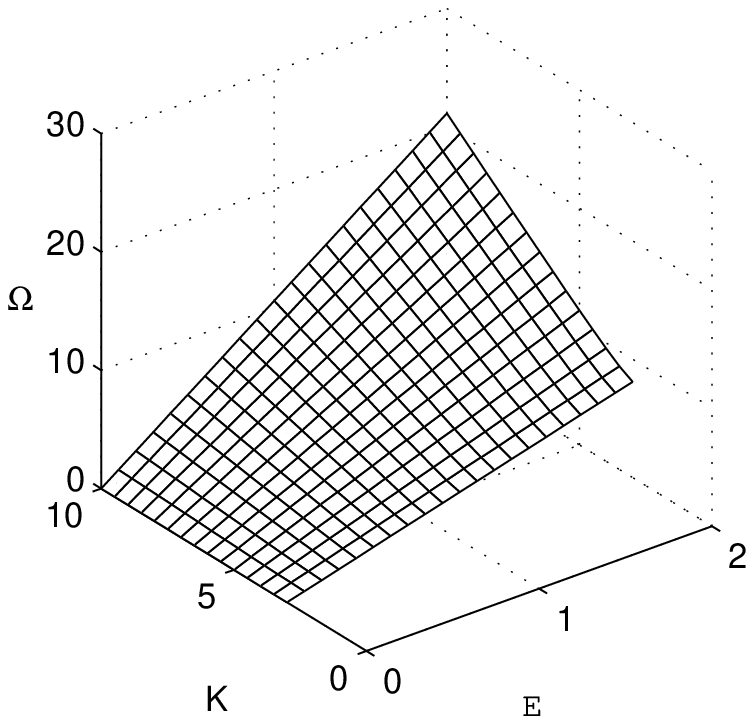} 
\caption{Dispersion surfaces $\Omega = \Omega(K,\mathcal{E})$ as given by
  Eq.\ (\ref{eq:norm}). The first panel corresponds to the $-$ sign in Eq.\ 
  (\ref{eq:norm}), and exists for all $K$ and $\mathcal{E}$. The second panel 
  shows the fast $+$ polarized mode, which exists for $K \gtrsim 2.6$. The third
  panel depicts the slow $+$ polarized mode, also for
  $K \gtrsim 2.6$.}
\end{figure*}

\end{document}